\documentclass[conference]{IEEEtran}
\IEEEoverridecommandlockouts

\usepackage{cite}
\usepackage{amsmath,amssymb,amsfonts}
\usepackage{algorithmic}
\usepackage{graphicx}
\usepackage{textcomp}
\def\BibTeX{{\rm B\kern-.05em{\sc i\kern-.025em b}\kern-.08em
    T\kern-.1667em\lower.7ex\hbox{E}\kern-.125emX}}
\usepackage[table]{xcolor}

\usepackage{acronym}
\usepackage{eqnarray}
\usepackage{arydshln}
\usepackage{booktabs}
\usepackage{hyperref}

\acrodef{ssl}[SSL]{Self Supervised Learning}
\acrodef{sl}[SL]{Supervised Learning}
\acrodef{byol}[BYOL]{Bootstrap Your Own Latent}
\acrodef{cnn}[CNN]{Convolution Neural Network}
\acrodef{ast}[AST]{Audio Spectrogram Transformer}
\acrodef{map}[mAP]{Mean Average Precision}
\acrodef{sota}[SOTA]{state-of-the-art}
\acrodef{vit}[ViT]{Vision Transformer}
\acrodef{ema}[EMA]{Exponential Moving Average}
\acrodef{mse}[MSE]{Mean Square Error}
\acrodef{mlm}[MLM]{Masked Language Modeling}
\acrodef{mlp}[MLP]{Multi Layer Perceptron}
\acrodef{met}[MATPAC]{\textbf{MA}sked laten\textbf{T} \textbf{P}rediction \textbf{A}nd \textbf{C}lassification}
\acrodef{ci}[CI]{Confidence Interval}
\acrodef{ml}[ML]{Multi-Label}
\acrodef{mcsl}[MCSL]{Multi-Class Single-Label}
\acrodef{mtt}[MTT]{Magna-Tag-A-Tune}
\acrodef{us8k}[US8K]{UrbanSound8K}

\newcommand{\pmval}[1]{{\fontsize{6pt}{6pt}\selectfont $\pm$ #1}}
\newcommand{\ie}{\textit{i.e.}}

\begin{document}

\title{Masked Latent Prediction and Classification for Self-Supervised Audio Representation Learning

\thanks{This work was supported by the Audible project, funded by French BPI, and was performed with GENCI-IDRIS ressources (Grant 2024-AD011013929R1)}
}

\author{\IEEEauthorblockN{Aurian Quelennec, Pierre Chouteau, Geoffroy Peeters, Slim Essid}
\IEEEauthorblockA{\textit{LTCI}, \textit{Télécom Paris}, \textit{Institut Polytechnique de Paris}, Palaiseau, France \\
    \{firstname\}.\{surname\}@telecom-paris.fr}
}

\maketitle

\begin{abstract}

Recently, self-supervised learning methods based on masked latent prediction have proven to encode input data into powerful representations. However, during training, the learned latent space can be further transformed to extract higher-level information that could be more suited for downstream \textit{classification} tasks. Therefore, we propose a new method: MAsked latenT Prediction And Classification (MATPAC), which is trained with two pretext tasks solved jointly. As in previous work, the first pretext task is a masked latent prediction task, ensuring a robust input representation in the latent space. The second one is unsupervised classification, which utilises the latent representations of the first pretext task to match probability distributions between a teacher and a student. We validate the MATPAC method by comparing it to other state-of-the-art proposals and conducting ablations studies. MATPAC reaches state-of-the-art self-supervised learning results on reference audio classification datasets such as OpenMIC, GTZAN, ESC-50 and US8K and outperforms comparable supervised methods' results for musical auto-tagging on Magna-tag-a-tune.

\end{abstract}


\section{Introduction}

Recently, \ac{ssl} methods have become a predominant representation learning approach across various application domains, especially speech and audio processing \cite{Liu22reviewaudio}. 
To alleviate the absence of labels in \ac{ssl}, one needs to elaborate pretext tasks where the model learns valuable representations from the input data itself. 
A wide variety of pretext tasks have been developed \cite{sslreview}. 
In general, models of this kind may operate either from a single view of the input \cite{huang2022MAE, assran23ijepa} or multiple views obtained through hand-crafted data augmentations \cite{Grill20byol, Niizumi23byola }. 
While the latter variant may be sensitive to the bias that results from the choice of views \cite{assran23vlusterprior}, single-view methods learn representations only from the original data, which requires less prior knowledge. 
One may also categorize pretext tasks with respect to the problem they solve, \ie, classification\cite{Caron21dino}, reconstruction \cite{huang2022MAE} or prediction \cite{assran23ijepa}. 

 \begin{figure}[htbp]
    \centerline{\includegraphics[width=0.9\linewidth]{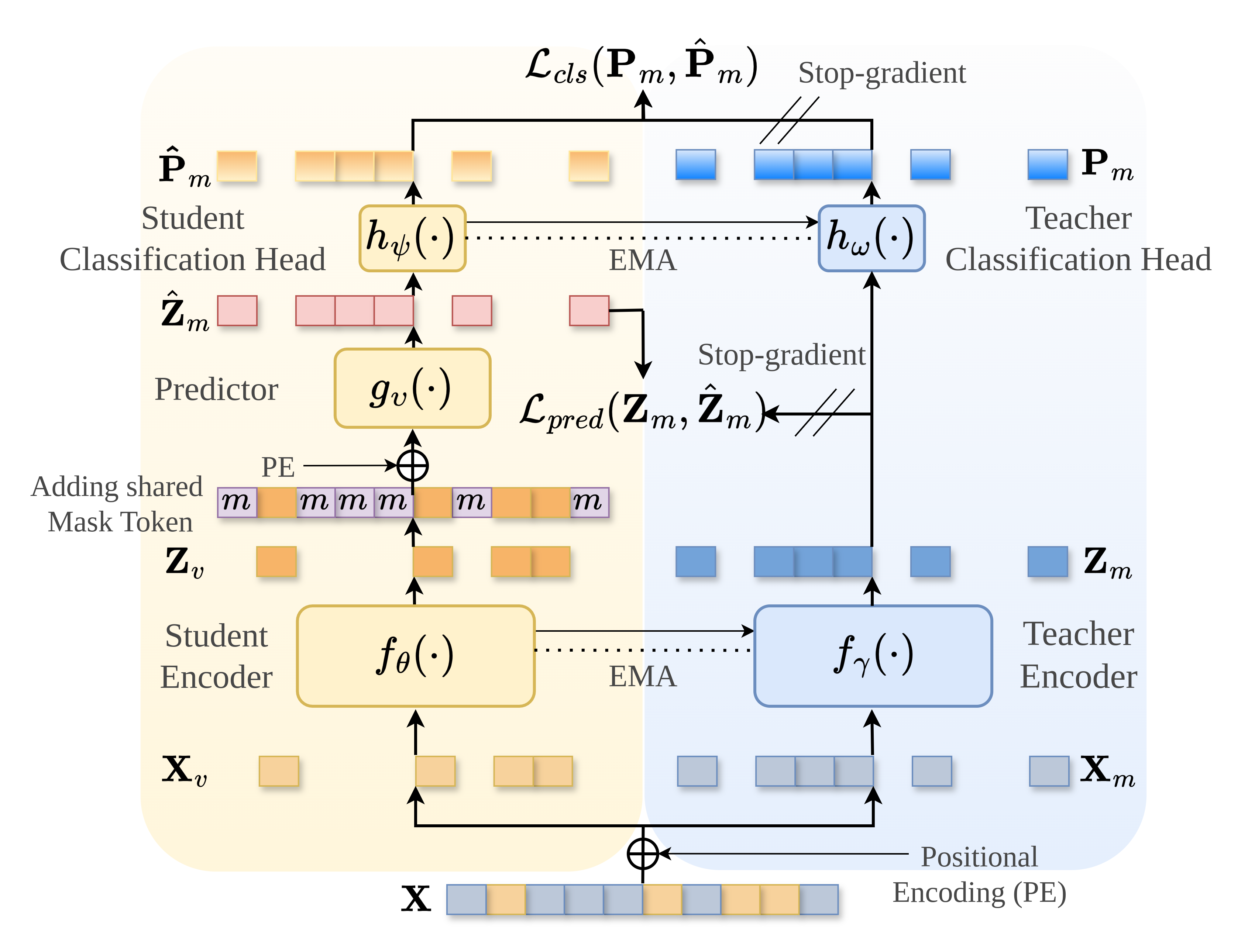}}
    \vspace{-0.2cm}
    \caption{Overview of proposed method, \ac{met}, combining a prediction and classification pretext task.} 
    \label{fig:main_pipe}
    \vspace{-0.6cm}
\end{figure}

Pretext tasks that consider unsupervised classification as an objective have proven quite successful. Models like HuBERT \cite{Hsu21HUBERT} or BEATs \cite{chen23beats} start by assigning  a cluster identity to each input, using either K-Means or a learned tokenizer, before learning a representation that can successfully predict the input--cluster association. In computer vision, DINO \cite{Caron21dino} tries to assign the same classes to multiple different views of the original input, using a teacher-student architecture.

More recently, promising results have been obtained using pretext tasks where the goal is to predict the latent representation of a masked part of the input from the visible part. 
This leverages the idea of \ac{mlm} \cite{Devlin19bert}.
When combined with a teacher-student paradigm, such as in \ac{byol} \cite{Grill20byol}, this leads to M2D \cite{niizumi23m2d} and I-JEPA \cite{assran23ijepa}.

\noindent\textbf{Contributions} \ \ 
This work stems from the intuition that in order to better target \textit{classification} downstream problems, an effective pretext task could be realised by modifying the goal of predicting the latent representation of a masked part of the input given a visible part (as done in I-JEPA and M2D), to instead predict the \textit{cluster identity} of the masked part. 

Thus, we propose a new \ac{ssl} method where we consider two pretext tasks which are solved jointly: \\
i) a \textit{masked prediction task}, solved in the latent space, as in previous work \cite{baevski22data2vec, niizumi23m2d, assran23ijepa}; \\
ii) an \textit{unsupervised classification task} where, as illustrated in Fig.\ref{fig:main_pipe}, the predicted and target representations are separately projected by a student and teacher classification head into probability distributions to be matched through a classification loss.

We call this new method: \acf{met}, which encodes the input into a robust predictive representation thanks to the first pretext task, while the classification pretext task makes it easier to extract high-level information (\textit{i.e.} cluster identity) further up from the latent representation, thus preparing the latent space for easy exploitation by classification downstream tasks. 

We show through extensive experimentation that \ac{met} sets new \ac{sota} performance for \ac{ssl} on musical instrument recognition, musical genre recognition and event classification tasks.


\section{Related Works}
\ac{ssl} methods based on masked prediction such as M2D \cite{niizumi23m2d} or I-JEPA \cite{assran23ijepa} rely on pretext tasks that exploit a single view of the original input, using a teacher-student architecture to solve a prediction problem in the latent space. 
The student encodes only the visible patches, while the teacher encodes the masked patches as a target latent representation. 
Then, from the latent representation of the visible patches, a shallow predictor tries to match the target latent representations.
While M2D was designed for audio signals and I-JEPA for images, they consider different masking strategies. 
Riou et al. \cite{riou24jepa} studied which of those masking strategies is the most suited for audio and found out that random masking is the most effective.

Besides, BEATs \cite{chen23beats} for general audio, HuBERT \cite{Hsu21HUBERT} for speech, or MERT \cite{Li24mert} for music, are models relying on unsupervised classification as a pretext task, but they either need iterative procedures or other models for initialization. 
Alternatively, other works in the field of computer vision use classification as an objective in their pretext task \cite{Caron21dino, Zhou22ibot, OquabA24dinov2}, considering multiple views of the input in a teacher-student architecture, thus avoiding the need for third-party models to provide the initial classification targets.
DINO \cite{Caron21dino} passes different augmentations of an input image to the student and teacher networks, projects them into a probability distribution, and learns to match them through classification.
In \cite{OquabA24dinov2}, the authors combine a DINO-like loss with the \ac{mlm} iBOT \cite{Zhou22ibot} loss to obtain a teacher-student system relying only on a classification objective. 

For learning general audio representations, SSAST \cite{Gong22ssast} and MAE-AST \cite{Baade22maeast} combine two pretext tasks: a masked input reconstruction and a classification objective. 
But as shown in \cite{niizumi23m2d}, using a masked prediction pretext task in the latent space is much more effective than reconstructing the masked part of the input. To our knowledge, no general self-supervised representation models combine a teacher-student architecture with both classification and prediction pretext tasks that consider the unsupervised classification of the latent representations resulting from a teacher-student prediction with masking.

\section{Method}

Fig. \ref{fig:main_pipe} depicts our method's entire pipeline. For the masked prediction pretext task, our approach is mainly inspired by M2D \cite{niizumi23m2d} and I-JEPA \cite{assran23ijepa} as we use a teacher-student architecture with a form of \ac{mlm} in the latent space.
In addition, we use the predicted and target latent representation to solve an unsupervised classification pretext task.

\noindent\textbf{Input of the model}\hspace{0.2cm} 
As input, we extract flattened patches, $\textbf{X}$, from the log-scale Mel spectrogram of each audio sample. 
A two-dimensional learned positional encoding, $\textbf{p}$ is added to $\textbf{X}$ before randomly partitioning the sequence into $\textbf{X}_v$, the visible patches, and $\textbf{X}_m$, the masked patches.
As opposed to I-JEPA \cite{assran23ijepa}, and like M2D \cite{niizumi23m2d}, we chose a random masking strategy.

\noindent\textbf{Masked prediction pretext task}\hspace{0.2cm}
The student encoder $f_{\theta}$, of parameters $\theta$, projects $\textbf{X}_v$ such that $\textbf{Z}_v = f_{\theta}(\textbf{X}_v)$ is the latent representation of the visible patches. 
Similarly, the masked patches are projected with the teacher encoder $f_{\gamma}$, and $\textbf{Z}_m = f_{\gamma}(\textbf{X}_m)$. The teacher's encoder parameters $\gamma$ are updated using an \ac{ema} of $f_{\theta}$. 
The update rule is $\gamma \xleftarrow{}\lambda\gamma + (1-\lambda)\theta$, were $\lambda$ is the decay rate. 

The input of the predictor, $g_{\upsilon}$, is all the encoded visible patches to which we append the same shared learnable token $\textit{\textbf{m}}$ at each masked position. 
A new learned positional embedding is added to give location information to the masked token. 
The predictor outputs a representation for visible and masked patch positions. 
However, we select only the prediction of the latent representation for masked patches to compare them with the targets.
The prediction for the masked latent representation is $ \hat{\textbf{Z}}_m = \{ g_{\upsilon}\left(\text{m\_concat}\left(\textbf{Z}_v, \textbf{m}, \mathcal{S}_m\right)\right)^{(i)}, i \in \mathcal{S}_m\}$, where $\mathcal{S}_m$ is the set of the masked indices. The prediction loss is then the square error of $l_2$-normalized versions of $\textbf{Z}_m$ and $\hat{\textbf{Z}}_m$, denoted by $\textbf{Z}'_m$ and $\hat{\textbf{Z}}'_m$:
\begin{equation}\label{eq:l_reg}
   \mathcal{L}_{pred}(\hat{\textbf{Z}}_m, \textbf{Z}_m) = \sum_{i}{ (\hat{\textbf{z}}'^{(i)}_m - \textbf{z}'^{(i)}_m )^2} \; ;
\end{equation}
where $\textbf{Z}'_m = \{\textbf{z}_m'^{(i)}, i \in \left[1,N\right]\}$, and $N$ is the number of targets, \textit{i.e.} the number of masked patches.

\noindent\textbf{Classification pretext task}\hspace{0.2cm}
Given a student $h_{\psi}$ and teacher $h_{\omega}$ projection heads, we transform the target and predicted latent representations into probabilities distribution of dimension $K$. 
$\textbf{P}_m$ and $\hat{\textbf{P}}_m$, respectively the target and predicted probability distributions, are obtained from $\textbf{Z}_m$ and $\hat{\textbf{Z}}_m$ as follow:
\begin{eqnarray}\label{eq:t_prob}
\hat{\textbf{P}}_m = \text{Softmax}( (h_{\psi}(\hat{\textbf{Z}}_m) / \tau_s); \\
\textbf{P}_m = \text{Softmax}( (h_{\omega}(\textbf{Z}_m) - \textbf{C}) / \tau_t);
\end{eqnarray}
where $\tau_s$ and $\tau_t$ are temperature parameters used to sharpen the distribution, and $\textbf{C}$ is used to center the distribution. 
$\textbf{C}$ is an \ac{ema} of the mean of $h_{\omega}(\textbf{Z}_m)$. 
Sharpening and centering were introduced by DINO's authors \cite{Caron21dino} to avoid the collapse of their method.
Our preliminary experiments showed that, even when the masked prediction pretext task is, we need the sharpening and centering operations to avoid the collapse of $h_{\omega}$ and $h_{\theta}$ into a trivial solution where one dimension dominates. 

Finally, the classification pretext task matches the distributions with a cross-entropy loss depending on $\psi$ of $h_{\psi}$: 
\begin{equation}\label{eq:losscls}
    \mathcal{L}_{cls}(\hat{\textbf{P}}_m , \textbf{P}_m)= \sum_{i}{-\textbf{p}^{(i)}_m\log(\hat{\textbf{p}}^{(i)}_m)}.
\end{equation}
The parameters $\omega$ of $h_{\omega}$ are updated with an \ac{ema} of $h_{\psi}$ with a decay rate $\zeta$ different from $\lambda$ (the decay rate of $f_{\theta}$ \ac{ema}).

\noindent\textbf{Total loss}\hspace{0.2cm} The final training objective is simply the sum of $\mathcal{L}_{cls}$ and $\mathcal{L}_{reg}$ with a weight $\alpha$:
\begin{equation}\label{eq:final_loss}
   \mathcal{L} = (1-\alpha)\mathcal{L}_{cls} + \alpha \mathcal{L}_{pred}.
\end{equation}

\section{Experiments}

In this section, we evaluate the \ac{met} method, showing 1) the effectiveness of combining masked prediction of the latent representation with unsupervised classification, 2) the balancing between the two tasks, and 3) the impact of the design of the classification head.

\subsection{Experimental Setup} \label{sec:exp_details}

\noindent\textbf{Pre-training dataset and input processing}\hspace{0.2cm} 
We use AudioSet \cite{gemmeke2017audioset} as a pre-training dataset. 
Our version of AudioSet has 2,012,215 samples over 6s length.
Each audio sample is processed to a log-scale Mel spectrogram with a sampling rate of $16,000$Hz using a window size of 25 ms, a hop size of 10 ms, and 80 Mel bins spaced between 50 and 8,000 Hz.
The log-scale Mel spectrograms are standardized with the dataset statistics. 
To process the log-scale Mel spectrogram into the input of the encoders, we used a patch size of $16 \times 16$ and a masking ratio of $0.7$, as it has shown to work well across different tasks \cite{niizumi23m2d, riou24jepa}.

\noindent\textbf{Encoders and predictor details}\hspace{0.2cm} 
We based our code on M2D's PyTorch implementation of the teacher-student encoder and predictor parts, making the minimum number of changes while keeping the same parameters. \footnote{\href{https://github.com/nttcslab/m2d}{https://github.com/nttcslab/m2d}}
To keep a fair comparison with \ac{sota} methods like M2D and ATST-Clip \cite{LiSL24atst}, and as it seems a good design choice \cite{riou24jepa}, we train our method with audio segments of 6 s. 
We randomly crop 6-s audio segments from the audio samples above this duration in the pre-training dataset.
Similarly to M2D, we train for 300 epochs, with a batch size of 2048, a warm-up of 20 epochs, the same base learning rate and optimizer, and the \ac{ema} decay rate $\lambda$ follows the same update policy.

\noindent\textbf{Classification head details}\hspace{0.2cm} 
We use the same architecture and parameters as in DINO for the teacher-student classification heads. \footnote{\href{https://github.com/facebookresearch/dino}{https://github.com/facebookresearch/dino}}
It is divided into two parts. 
First, there are three fully connected layers with a bottleneck output of dimension 256. 
Then, a weight-normalized fully connected layer projects the $l_2$-normalized output of the first layers into the probability distribution space of $K$ dimensions~\cite{Caron21dino}.
By default, we set the hidden dimension to 2048 and $K=2048$.
We linearly increase the temperature parameter $\tau_t$ in Eq. \eqref{eq:t_prob}, from 0.04 to 0.07 for $n_{\tau,epoch}$ epochs. 
This encourages the target's probability distribution to be peakier at the beginning of the training, forcing some ``classes" to appear.  
We keep $\tau_s=0.1$ for the whole training.
Finally, the \ac{ema} decay rate $\zeta$ of the classification heads update is linearly interpolated from 0.998 to 1 at the end of the training and $\alpha=0.5$. 

\begin{table}[htb!]
    \caption{Evaluation Downstream tasks. TVT stands for Train/Validation/Test}
    \centering    
    \resizebox{0.75\columnwidth}{!}{
    \begin{tabular}{c| c c c c c}
        Dataset & \#Samples & \#Classes & Type & Split & Metric  \\
        \midrule
        OpenMIC & 20,000 & 20 & ML & TVT  & mAP \\
        NSynth & 305,979 & 11 & MCSL & TVT & Acc. \\
        GTZAN & 930 & 10 & MCSL & TVT & Acc. \\
        MTT & 25,863 & 50 & ML & TVT & mAP \\
        FSD50K & 51,197 & 200 & ML & TVT & mAP \\
        ESC-50 & 2,000 & 50 & MCSL & 5-Fold & Acc. \\
        US8K & 8,732 & 10 & MCSL & 10-Fold & Acc. \\
    \end{tabular}
    }
    \vspace{-0.6cm}
    \label{tab:datasets}
\end{table}

\noindent\textbf{Linear probing evaluation}\hspace{0.2cm}
We train a linear classifier consisting of a single fully connected layer that maps the latent representation of the model we evaluate to the number of classes of the task.
We use a batch size of 128 for all tasks and a learning rate of $1e^{-4}$ with an ADAM optimizer. 
We kept the model with the best validation loss, and each downstream experiment was repeated five times. 
We report the average scores with their \ac{ci} at $95\%$.

For \ac{met}, if the input audio from the downstream task exceeds 6s, we split the audio every 6s without overlapping and perform average pooling over time to have one embedding per clip. Since our method encodes each time-frequency patch (extracted from the input log-scaled Mel spectrogram) with a specific latent representation, we concatenate the patches over frequencies and average them over time, as done in M2D.

\begin{table*}[htb!]
    \footnotesize
    \caption{Linear evaluation comparison with \ac{ci} at 95\%. $\dagger$ scores and \ac{ci} are from the original paper. \ac{sl} methods are greyed out. Bold scores are the best \ac{ssl} results. Underlined ones are the best scores overall. Possible training data are AudioSet (AS) and LibriSpeech (LS). }
    \centering
    \begin{tabular}{ l l l l l l l l l l }
              & & OpenMIC & NSynth & GTZAN & MTT & FSD50K & ESC-50 & US8K & Avg. \\
        Models & Training data & mAP & Acc(\%) & Acc(\%) & mAP & mAP & Acc(\%) & Acc(\%) &  \\
        
        \midrule
        \ac{met} - $n_{\tau,epoch}=10$ & AS & 85.3\pmval{0.0} & 74.3\pmval{0.2} & 85.3\pmval{0.4} & \underline{\textbf{41.1}}\pmval{0.0} & 55.2\pmval{0.1} & \textbf{93.5}\pmval{0.1} & \textbf{89.4}\pmval{0.1} & \textbf{74.9} \\
        
        \ac{met} - $n_{\tau,epoch}=20$ & AS & \textbf{85.4}\pmval{0.0} & 74.6\pmval{0.5} & \textbf{85.9}\pmval{0.3} & 41.0\pmval{0.0} & 54.8\pmval{0.1} & 93.2\pmval{0.1} & 88.0\pmval{0.0} & \textbf{74.7} \\

        \addlinespace[-0.01cm] \hdashline \addlinespace[0.05cm]
        
        \ac{met} w/o Classification& AS & 85.1\pmval{0.0} & 73.7\pmval{1.1} & 84.2\pmval{0.4} & 40.8\pmval{0.0} & 54.4\pmval{0.2} & 92.5\pmval{0.1} & 87.5\pmval{0.1} & 74.0  \\ 
        
        \midrule
        M2D - $\text{ratio}=0.7$ \cite{niizumi23m2d} & AS & 84.8\pmval{0.1} & $\underline{\textbf{76.9}}^{\dagger}$\pmval{1.3} & $83.9^{\dagger}$\pmval{1.4} & 40.6\pmval{0.1} & 52.8\pmval{0.8} & $89.8^{\dagger}$\pmval{0.3} & $87.1^{\dagger}$\pmval{0.3} & 73.7\\
        
        ATST-Clip \cite{LiSL24atst} & AS &  84.2\pmval{0.0} & $76.2^{\dagger}$ & 79.9\pmval{0.2} & 39.1\pmval{0.0} & $\textbf{58.5}^{\dagger}$ & 91.4\pmval{0.1} & $85.8^{\dagger}$ & 73.6 \\
        
        ATST-Frame \cite{LiSL24atst}& AS & 83.1\pmval{0.1} & $75.9^{\dagger}$ & 80.7\pmval{0.7} & 39.1\pmval{0.1} & $55.1^{\dagger}$ & 87.5\pmval{0.2} & $85.8^{\dagger}$ & 72.5 \\
        
        BEATs iter3 \cite{chen23beats}& AS  & 82.7\pmval{0.1} & 72.4\pmval{0.7} & 80.0\pmval{0.3} & 38.3\pmval{0.0} & 46.7\pmval{0.1}& 87.5\pmval{0.1} & 85.4\pmval{0.1} & 70.4 \\
        
        MAE-AST \cite{Baade22maeast}& AS+LS & 79.4\pmval{0.1} & 71.2\pmval{0.4} & 64.1\pmval{1.2} & 36.9\pmval{0.2} & 41.1\pmval{0.1} & 84.4\pmval{0.2} & 81.3\pmval{0.2} & 65.5\\
        
        \midrule
        
        \textcolor{gray}{PaSST} \textcolor{gray}{\cite{Koutini22passt}}& AS & \textcolor{gray}{86.5\pmval{0.1}} & \textcolor{gray}{72.9\pmval{0.5}} & \textcolor{gray}{\underline{87.4}\pmval{2.0}} & \textcolor{gray}{40.4\pmval{0.1}} & \textcolor{gray}{\underline{61.3}\pmval{0.2}} & \textcolor{gray}{\underline{97.0}\pmval{0.1}} & \textcolor{gray}{89.1\pmval{0.1}}  & \textcolor{gray}{76.4}\\
        
        \textcolor{gray}{HTS-AT}\textcolor{gray}{\cite{Chen22htsat}} & AS & \textcolor{gray}{86.4\pmval{0.0}} & \textcolor{gray}{68.6\pmval{0.3}} & \textcolor{gray}{85.9\pmval{0.4}} & \textcolor{gray}{40.1\pmval{0.0}} & \textcolor{gray}{59.4\pmval{0.0}} & \textcolor{gray}{95.9\pmval{0.1}} & \textcolor{gray}{85.3\pmval{0.0}} & \textcolor{gray}{74.5} \\
        
        \textcolor{gray}{BEATs iter3+} \textcolor{gray}{\cite{chen23beats}}& AS & \textcolor{gray}{\underline{86.7}\pmval{0.0}} & \textcolor{gray}{71.6\pmval{0.5}} & \textcolor{gray}{86.0\pmval{0.4}} & \textcolor{gray}{40.3\pmval{0.0}} & \textcolor{gray}{60.6\pmval{0.1}} & \textcolor{gray}{96.1\pmval{0.1}} & \textcolor{gray}{\underline{89.5}\pmval{0.1}} & \textcolor{gray}{75.8}\\ \\

\end{tabular}
\vspace{-0.6cm}
    \label{tab:gen_res}
\end{table*}

\noindent\textbf{Downstream Datasets}\hspace{0.2cm} 
To evaluate our method, we selected a variety of classification downstream tasks that cover music and environmental sound domains. 
For instrument classification tasks, we have chosen the NSynth dataset \cite{EngelRRDNES17nsynth}, a synthetic dataset of musical notes.
We also consider OpenMIC \cite{Humphrey18openmic} for instrument classification as its samples are extracted from real music recordings.
For genre classification tasks, we use GTZAN \cite{TzanetakisC02gtzan} with the corrected labels following Sturm's work \cite{Sturm13gtzancorr}. 
For music autot-agging, we use the \ac{mtt} dataset.
While GTZAN and NSynth are widely used in evaluation frameworks \cite{turian2022hear, lWang2022hares, Niizumi23byola}, OpenMIC and \ac{mtt} are less common.
For environmental sound analysis tasks, we consider FSD50K \cite{FonsecaFPFS22fsd50k}, a sound event event recognition dataset whose classes are from the AudioSet ontology.
In addition, we consider the widely-used ESC-50 \cite{Piczak15esc50}, and \ac{us8k} \cite{SalamonJB14us8k} which is an urban sound classification dataset. 
Table \ref{tab:datasets} provides more details about each of the datasets, for instance, the type of classification, which is either \ac{ml} or \ac{mcsl}.

\subsection{Comparison with \ac{sota} methods}\label{sec:effectiveness_method}
We compare \ac{met} with other \ac{sota} methods on linear evaluation. Results shown in Table \ref{tab:gen_res} prove that our method is better than previous \ac{ssl} methods on all tasks except NSYNTH and FSD50K, where respectively M2D and ATST models get higher scores.
Performance discrepancy between M2D and \ac{met} without the classification task are mostly due to the differences in the list of AudioSet excerpts used in our experiment (owing to no longer active YouTube videos).
Moreover, while on most tasks, the supervised learning methods PaSST \cite{Koutini22passt}, HTS-AT \cite{Chen22htsat} and BEATs iter3+ still outperform all self-supervised methods, \ac{met} achieves score of 41.1\% on \ac{mtt}, outperforming both supervised and self-supervised learning methods considered here\footnote{Note that \ac{sota} scores on MTT are achieved by \cite{CastellonDL21mttsota}}. 
\ac{met} also has comparable performances with supervised methods for instrument classification on GTZAN and urban sound classification on US8K.

The code and pre-trained weights are publicly available.\footnote{\href{https://github.com/aurianworld/matpac}{https://github.com/aurianworld/matpac}}

\subsection{Ablation experiments}
\begin{table}[htb!]
    \footnotesize
    \caption{Impact of $\alpha$, the weighting  between the classification and the masked prediction pretext task}
    \centering
    \begin{tabular}{c | c c  c c}
        $\alpha$ & 0.25 & 0.5 & 0.75 & 1    \\
        \midrule
        
        Music & \textbf{71.5} & \textbf{ 71.5} & 71.4 &  71.0 \\
        
        Environment & 78.5  &\textbf{ 79.3} & 78.5 & 78.1 \\
        
        All & 74.5 &\textbf{ 74.9} & 74.4 & 74.0  \\
        
    \end{tabular}
    \label{tab:alpha}
    \vspace{-0.2cm}
\end{table}

\noindent\textbf{Importance of solving jointly the masked prediction and classification pretext tasks}\hspace{0.2cm}\label{sec:loss_balance}
Firstly, in Table \ref{tab:gen_res}, we observe that, for both $n_{\tau,epoch}$ values, \ac{met} has better scores to those without the classification pretext task.
The two configurations achieved an average score of 74.9 and 74.7, significantly above the average score of 74.0 obtained when training without the classification pretext task.
For $n_{\tau,epoch}=20$ epochs, \ac{met} achieves the best overall score on OpenMIC, NSYNTH and GTZAN, while it has better scores on environmental tasks for $n_{\tau,epoch}=10$.
Without the masked prediction pretext task, however, the model collapses as the classification head is not able to ensure that $\textbf{Z}_m$ is close to $\hat{\textbf{Z}}_m$.
Therefore $\textbf{P}_m$ and $\hat{\textbf{P}}_m$ do not match either. 
Those results emphasise the utility of considering both pretext tasks: masked prediction and unsupervised classification.

\noindent\textbf{Influence of $\alpha$ in the loss}\hspace{0.2cm}\label{sec:loss_balance}
In Table \ref{tab:alpha}, we show that $\alpha=0.5$ yields the best average score of 74.9. For lower values, the classification pretext task has more importance in the loss and scores start to drop. On the other hand, when $\alpha=1$, only the masked prediction pretext task is taken into account in the loss Eq. \eqref{eq:final_loss}, and the average score of 74.0 shows that it is less effective than using both pretext tasks.

 \begin{figure}[htbp]
    \centerline{\includegraphics[width=0.85\linewidth]{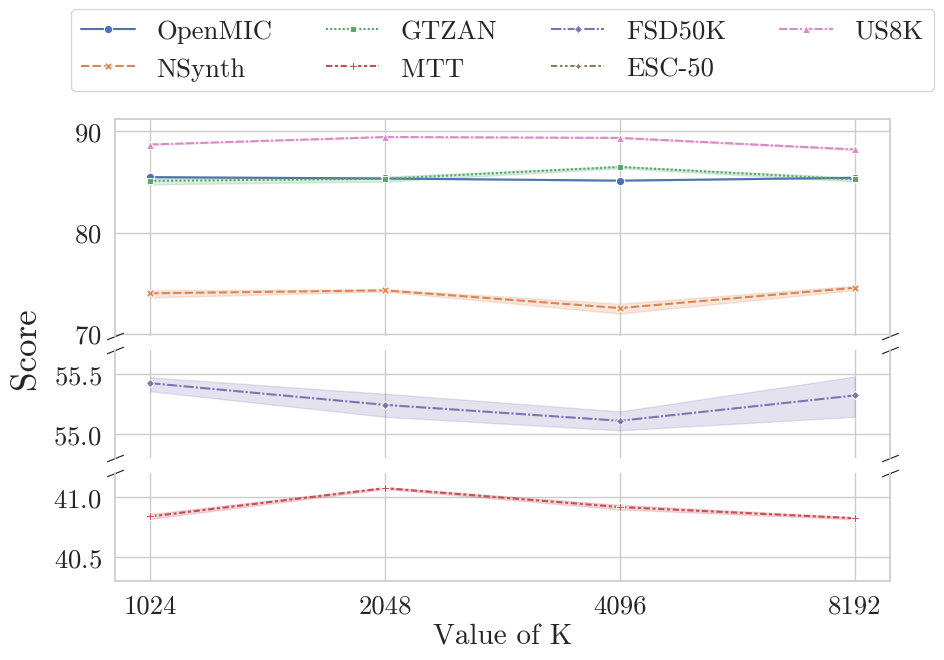}}
    \vspace{-0.3cm}
    \caption{Influence of the dimensionality $K$ in the classification head} 
    \label{fig:headdim}
    \vspace{-0.4cm}
\end{figure}

\noindent\textbf{Classification head architecture}\hspace{0.2cm}
Concerning the classification head, we explore which output dimensionality $K$ is best. 
In Fig. \ref{fig:headdim}, we report the scores on each downstream task for $K \in [1024,2048,4096,8192]$. 
We observed that scores vary very little as a function of $K$, but overall, for $K=2048$ we obtain the best results. 
We can conclude that the classification pretext task only plays the role of regularizer of the latent space, making the representation effective in downstream tasks. 

\section{Conclusions}

We proposed a new Self Supervised Learning method, \acf{met}, which combines two pretext tasks to learn a robust representation for classification downstream tasks. 
The first pretext task performs masked prediction in the latent space, which had already proved effective in previous work. 
The second is an unsupervised classification task that projects the first task's predicted and target latent representation into probability distributions matched through a classification loss. 
Through our evaluation of \ac{met} on various downstream tasks and our ablation studies, we prove the effectiveness of solving two pretext tasks jointly.
Notably, \ac{met} achieves state-of-the-art results compared to other \ac{ssl} methods on OpenMIC, GTZAN, ESC-50 and US8K while outperforming comparable fully supervised competitors on NSynth and Magna-tag-a-tune.

\bibliographystyle{IEEEtran.bst}
\bibliography{biblio}

\end{document}